\documentclass[aps,pre,preprint,superscriptaddress,amsmath,amssymb,thelongbibliography,nofootinbib,showpacs]{revtex4-1}
\usepackage{amsmath} 
\usepackage{graphicx} 
\usepackage{amssymb} 
\usepackage{mathtools}

\usepackage[final]{hyperref}
\hypersetup{backref,
colorlinks=true}
\usepackage{dcolumn} 
\usepackage{bm} 
\usepackage{color} 
\usepackage{float} 
\usepackage{upgreek} 
\usepackage[normalem]{ulem}

\makeatletter 

\begin{document} 

\title{Effective medium model for a suspension of active swimmers} 

\author{A. Dhar}\thanks{abyaya93@iitkgp.ac.in} 
\affiliation{Department of Physics, Indian Institute of Technology, Kharagpur, 721302, India} 

\author{P.S. Burada}\thanks{psburada@phy.iitkgp.ac.in} 
\affiliation{Department of Physics, Indian Institute of Technology, Kharagpur, 721302, India} 

\author{G.P. Raja Sekhar}\thanks{rajas@iitkgp.ac.in} 
\affiliation{Department of Mathematics, Indian Institute of Technology, Kharagpur, 721302, India} 

\date{\today} 

\begin{abstract}
Several active organisms in nature tend to reside as a community in a viscous fluid medium. We analyze the variation of swimming characteristics of an active swimmer present in a dilute and disperse suspension, modeled as an effective Brinkman medium. This idealized representation of a collection of active swimmers allows one to distinguish the impact of the interior domain available to an individual swimmer as well as the contribution of its neighbours. The Darcy's law along with the analytical solution enable the effective resistivity to be predicted as a function of the volume fraction which is in close agreement with the well known Carman-Kozeny equation. This facilitates the successive analysis of the propulsion speed, power dissipation, and swimming efficiency of the targeted swimmer as a function of the volume fraction which are decisive in nutrient transport and uptake or reproduction in a collective environment. A stress-jump condition is also imposed across a cell to indicate a mean effective force due to the nearby swimmers. At suitable values of this stress-jump coefficient, the relative increase of the migration velocity and swimming efficiency is noticeably higher at an optimum occupancy.         
\end{abstract}
\maketitle

\section{Introduction}
\label{Intro}

A collection of active systems exhibits a wide range of behaviors that emerge due to their intrinsic active constituents. Intriguing behaviors are observed in systems ranging from the macroscopic scale of flocks of birds \cite{Cavagna}, schooling of fish \cite{Pavlov}, and mammalian herds \cite{Brock} to the microscopic scale of bacterial suspensions and swarming \cite{Copeland}. Properties of these strongly self-organizing dissipative systems are of fundamental interest to encompass applications in non-equilibrium statistical mechanics \cite{Vicsek, Czirokdag, Simha}, nonlinear dynamics \cite{Saintillan, Dhar-JFM}, and microfluidic technologies \cite{Ebbens, Nelson}. 

Most of the real-life active microorganisms such as {\it Volvox}, {\it Paramecium} and many others propel through their corresponding propulsive appendages, i.e., flagella or cilia, at the smallest scales of their movement \cite{Shum}. Migration of such organisms is aptly described by the classical `squirmer' model introduced by Lighthill \cite{Lighthill} and furthered by Blake \cite{Blake}. The asymmetric movement of these propulsive appendages is modeled as a slip velocity between the squirmer's surface and the suspending fluid. With the most straightforward possible description, the squirmer model has successively explained the behavior exhibited by active organisms in an unbounded fluid \cite{Pedley, Drescher}, the hydrodynamic interaction amidst them \cite{Ishikawa-JFM, AVSingh}, and the interaction with a wall \cite{JSL,Jalilvand}. Extensive hydrodynamic analysis of the squirmer model \cite{Lauga-powers, Dhar-IJMF} have upgraded our general understanding of low-Reynolds-number locomotion, which has also incited the design of artificial microswimmers \cite{Dhar-2019, Dhar-2020, Haan}. Recent developments have proposed self-propulsion methods, at the micro-scale, using the artificial appendages \cite{Ghanbari,Zhang}, commencing the prospects of experimental realization of the squirmer slip model. For application in experimental studies, another particularly appealing strategy is to induce a phoretic slip on the surface of chemically active colloids \cite{MNPopescu} where the induced diffusiophoretic slip due to an inhomogeneous chemical field can be regarded to a structure analogous to the squirmer slip for future analysis. 

Therefore, the suspension of such squirmers is a convenient representation of an active ensemble. In the past, for self-organizing active systems, the interactions are incorporated in various means \cite{Ishikawa-JFM, Ishikawa-2, Zhang-int, Lega}. One may note that, in the case of suspension of squirmers, the hydrodynamic interactions are eminent in case of dense suspensions and establish a long-range order in the system \cite{Dombrowski}. Subsequently, the self-organized coherent structures, exceeding the size of an individual swimmer, start occurring \cite{Sokolov, Hill}. Contrary to this, the hydrodynamic interactions are often ignored in dilute and intermediate suspensions as the interparticle distance exceeds the range of hydrodynamic interaction \cite{Haines}. Moreover, in the case of disperse suspensions, it is viable to characterize one single squirmer and represent the nearby interactions as an average or effective force over a range which gives rise to a unique environment comprising small compact structures at a given instant of time. 

The effective Brinkman medium via the concept of cell model \cite{Li-int, Dodd} adequately represents this interspersed arrangement of a disperse suspension, which allows one to identify both the environment around the individual particle as well as the interactive force. In the case of impenetrable or solid systems, the model has been applied to a granular media comprising colloidal spheres, which have shown good agreement with the existing numerical models \cite{Li-int}. Besides that, Dodd et al.\cite{Dodd} have successfully used the effective medium approximation to understand the hydrodynamic interaction on the diffusivities of integral protein membranes. A primary concern in analyzing a suspension of active systems is the distribution of squirmers in the domain, which plays a crucial role in determining the intermediate interactions \cite{Ishikawa-4,Gyrya}. The effective medium approximates \cite{Li-int} the active suspension as the collection of several cells, a fictitious fluid envelope, spherical or cylindrical in shape \cite{Biesheuvel}, formed concerning the squirmers embedded in the medium \cite{Happel-eff}. The entire disturbance generated due to each particle is confined to the cell associated with it, i.e., a fluid envelope surrounding it. Further, it is assumed that the impact of neighbouring swimmers are confined to the fluid envelope only and can be incorporated through suitable boundary conditions across the cell envelope, i.e., the fluid-porous interface of the system \cite{JPrakash}. It is then theoretically possible to obtain a closed solution which later prompts a concentration dependence. In the present work, we assume a stress jump condition, proposed by Ochoa-Tapia and Whitaker \cite{Ochoa-Tapia-1,Ochoa-Tapia-2}, to model the mean interactive force across the cell developed due to nearby swimmers. Earlier, the stress jump condition has been widely used in various models \cite{JPrakash, Bhattacharya} in order to realize its significance in various physical situations. For a cell model like homogeneous suspension, the nearby squirmers with respect to the targeted squirmer are at an equivalent physical distance which justifies the simplifying assumption that the cell envelope, encapsulating a single squirmer, encounters a mean isotropic stress jump due to the presence of immediate neighbours in the vicinity.  

In this work, we aim to highlight the significant changes in individual and collective responses of a suspension of squirmers within this effective medium and cell model realm. Most of the existing works on active suspensions neglect the compact microstructures in the suspension. These fundamental structures play an important role to predict the particle removal rate in the fibrous and granular media \cite{Tien}. Towards this, the present investigation constitutes a systematic step in understanding and utilizing the active suspension's novel properties. Additionally, it brings out the significance of the interactive stress jump condition in governing the swimming properties. One may note that understanding individual squirming properties in the presence of multiple squirmers is still developing, and a robust numerical simulation at a micro-scale may be authoritative. However, the individual interaction through Stokes' simulation or an approximated Langevin simulation will either claim extensive computation resources or questionable accuracy. The present analytical framework of the cell model or the effective medium model provides a rough estimate in obtaining the swimming characteristics that qualitatively captures the empirical situation. 

\section{Mathematical formulation}
\label{math-model}

\begin{figure}
\includegraphics[scale=0.3]{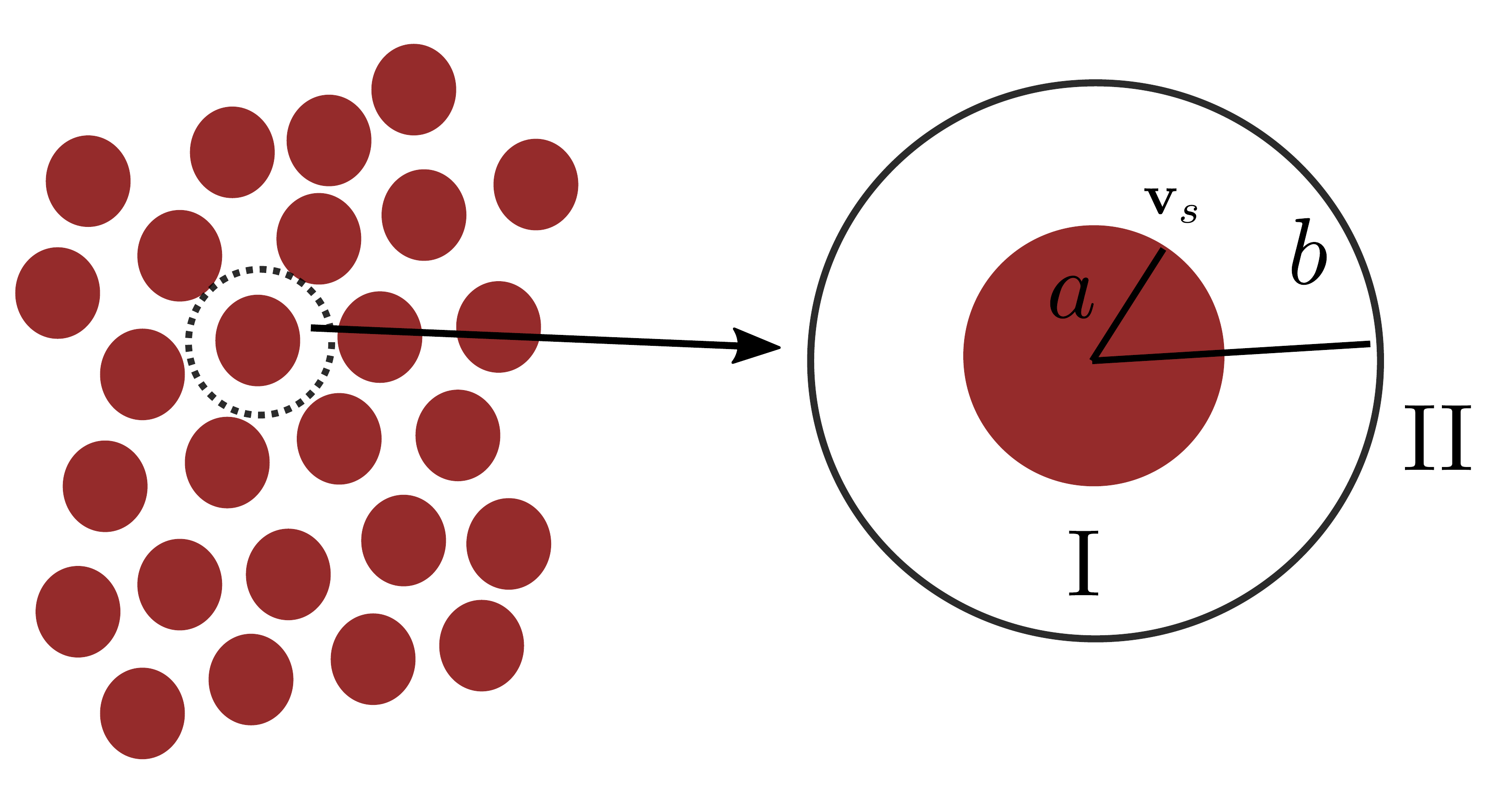}
\caption{\label{Fig1} Geometry of the effective Brinkman model comprising identical squirmers of radius $a$ and prescribed tangential slip $\mathbf{v}_s$. The cell's position $b$ fixes the interaction range between the squirmers and subsequently determines the effective Brinkman medium's packing fraction. The region {\bf I} indicates the Stokesian regime available to individual squirmers, whereas the region beyond the cell, i.e., {\bf II} marks the effective Brinkman domain.}
\end{figure}

We consider a homogeneous collection of spherical squirmers each with radius $a$ and prescribed tangential slip velocity $\mathbf{v}_s$. At the micro-scale, each one of the squirmers is surrounded by a hypothetical fluid envelope of radius $b$. Such envelope is referred to as Chang’s unit cell \cite{Chang}. On the other hand, at the macro scale, we treat this collection as a bed of porous medium. More recently, Nganguia et al. \cite{Nganguia-2} have employed a similar mathematical structure via squirmer model to probe swimming of a $H. pylori$ bacterium which creates and modifies a biological barrier during migration. Here, we assume a similar structure for a suspension containing sufficiently many squirmers such that beyond the individual cellular space, the medium can be treated as an equivalent Brinkman medium \cite{Li-int} so as to produce appreciable changes in the properties of any constituent squirmer. As on a macroscopic scale the suspension adheres to the Darcy's law, an interacting suspension of force free squirmers can be hold together as a porous medium only in presence of a constant external macroscopic pressure gradient $\nabla p^f$ \cite{Carman-Book}. We also assume that the disturbance flow produced by a single squirmer is negligible at the locations of other squirmers, consequently, we avoid incorporating explicit hydrodynamic interactions among the squirmers, and the intermediate interactions of the neighbors are only restricted to the cell \cite{Happel-eff,Kuwabara,Kvashnin} that are taken into account by the suitable boundary conditions across the cell boundary \citep{Dodd}. We chose the size of the cell such that the volume fraction of the dispersed phase in each cell matches with the overall mean volume fraction of the dispersed phase in the ensemble, i.e., the porosity of the effective Brinkman model $\Xi$ \cite{Davis}. The fluid element within the cell remains as a purely viscous medium which further permits the consideration of Stokes’ flow due to the negligible Reynolds number limit for the squirmers. Whereas, we use the incompressible Brinkman equation to govern the resistive network formed by the neighboring squirmers. Subsequently, we have
\begin{align}
\mu \nabla^2 \mathbf{v}^{en} - \nabla p^{en}=0, \quad \nabla \cdot \mathbf{v}^{en}=0 \quad a<r<b, \\
\mu \nabla^2 \mathbf{v}^e - \frac{\mu}{k^e} \mathbf{v}^e -\nabla p^{e}=0, \quad \nabla \cdot \mathbf{v}^e=0 \quad r>b,
\end{align}
where $(\mathbf{v}^{en}, p^{en})$ and $(\mathbf{v}^e, p^e)$ represent the velocity and pressure fields interior and exterior to the cell envelope, respectively. The momentum equation for $r>b$ combines both the effect of viscous stresses and the effect of a distributed Darcy force which is characterized by the Darcy permeability $k^e$ that depends on the structure of the porous medium formed by the suspension of the squirmers. The governing equations are asserted in the spherical polar coordinate ($r,\theta,\phi$) systems. Subsequently, we introduce the dimensionless variables as $\tilde{\mathbf{r}}=\mathbf{r}/R$, $\tilde{\mathbf{v}}=\mathbf{v}/U$, and $\tilde{P}=P/\mu\, U/a$, where $\mu$ is the viscosity of the suspending medium and $U=\mid \mathbf{U} \mid$ is the magnitude of the characteristic far-field migration velocity of any squirmer packed in the ensemble. Further, we have dropped the tilde notation for convenience. Therefore, the system of equations in the non-dimensional form is given by
\begin{align}
\nabla^2 \mathbf{v}^{en} & = \nabla p^{en}, \quad \nabla \cdot \mathbf{v}^{en}=0, \quad 1<r<b/a=j, \label{gov-1-ch7} \\
\left( \nabla^2 - \chi_e^2 \right) \mathbf{v}^e & = \nabla p^e, \quad \nabla \cdot \mathbf{v}^e=0, \quad r>j \label{gov-2-ch7},
\end{align}
where $\chi_e= a^2/k_e$ is the dimensionless resistance of the medium which compares the radius of the squirmer $a$ to the permeability $k^e$ of the effective Brinkman medium. As $\chi_e$ approaches zero, the effective Brinkman equation reduces to the Stokes equation which is equivalent to the case of a single squirmer in an unbounded medium and for large $\chi_e$, the equation reduces to the Darcy equation. Here, the non-dimensional position of the cell $j=b/a$, the volume fraction of the squirmers $\Phi$, and the porosity of the effective Brinkman medium are related as $\Phi=(1-\Xi)=1/j^3$. It is ingrained to assume that the inverse permeability ($\chi_e$) depends on porosity or the volume fraction ($\Phi$) of the medium and one of the most widely accepted theories for the permeability-porosity relationship is the Kozeny-Carman (KC) model \cite{Carman-Book} which provides a link between medium properties and flow resistance in pore channels. However, in the present work, the effective porous medium constitutes dynamic squirmers, and subsequently, the skeleton of the effective porous medium is not defined beforehand. Therefore, it is not straightforward to determine the appropriate relationship as this would require an elaborate knowledge of the distribution and spatial arrangement of the squirmers forging the effective porous medium. However, later in this work, after acquiring the analytical solution we will attempt to establish an analogous empirical relation for an active suspension of squirmers.

A typical squirmer model assumes an active slip over the entire surface \cite{Lighthill, Blake}, consequently, at the surface of the squirmer ($r=1$), the normal component of the velocity field vanishes whereas the tangential components agree to the squirmer slip distribution, as
\begin{align}
\mathbf{v}^{en}\cdot \mathbf{\hat{n}} & = 0, \label{bc-1-ch7} \\
\mathbf{v}^{en} \cdot \mathbf{\hat{t}} & = \mathbf{v}_s \cdot \mathbf{\hat{t}} \label{bc-2-ch7},
\end{align}
where $\mathbf{v}_s = \beta\, \nabla_s P_1(\cos\theta)$ is the active slip decomposed in terms of a Legendre polynomial of order one $P_1(\cos\theta)$ with $\beta$ denoting the strength and $\theta$ being the acquired polar angle from the symmetry axis. $\mathbf{\hat{n}}$ and $\mathbf{\hat{t}}$ are the relevant normal and tangential unit vectors in the spherical polar coordinate system. For squirming motion in Stokes flow, $\beta$ corresponds to a source dipole and is the only mode contributing to propulsion speed. Here, we stick to the first mode of the slip velocity $\mathbf{v}_s$ and ignore the higher modes in order to study the migration of a squirmer assemblage. Further, appropriate attention must be given to the conditions at the fluid-porous interface across the cell boundary, i.e, at $r=b/a=j$, which plays a crucial role in determining the interactions among the pack of the squirmers suspended in the medium. The standard conditions at porous-fluid interface depend on the nature of porous material and the corresponding governing equations that are employed inside the porous medium \citep{Beavers, Nield}. However, in the case of Stokes-Brinkman coupling, one may use the continuity of velocity and normal stress components together with a jump in tangential stress which is popularly known as the stress jump condition \citep{Alberto-1,Alberto-2,JPrakash}. Accordingly, we have (at $r=j$)

\begin{enumerate}
\item[$\bullet$] continuity of the velocity components:
\begin{equation}
\mathbf{v}^{en}=\mathbf{v}^{e} \label{bc-3-ch7},
\end{equation}

\item[$\bullet$] continuity of the normal stress component:
\begin{equation}
\mathbf{\Sigma}_{nn}^{en}=\mathbf{\Sigma}_{nn}^{e} \label{bc-4-ch7},
\end{equation}

\item[$\bullet$] stress-jump boundary condition for the shear stress:
\begin{equation}
\frac{\partial (\mathbf{v}^e \cdot \mathbf{\hat{t}})}{\partial n} - \frac{\partial (\mathbf{v}^{en} \cdot \mathbf{\hat{t}})}{\partial n} = \alpha_e \chi_e \left( \mathbf{v}^e \cdot \mathbf{\hat{t}} \right) \label{bc-5-ch7},
\end{equation}
\end{enumerate}

where $\mathbf{\Sigma}^{e}$ and $\mathbf{\Sigma}^{en}$ stress tensors corresponding to the fluids exterior and interior to the cell envelope, respectively. Here, $\alpha_e$ indicates the stress jump coefficient which entails the effective interactive stress across the cell boundary developed due to the existence of identical squirmers in its vicinity. The founding work by Ochoa-Tapia and Whitaker \cite{Ochoa-Tapia-1, Ochoa-Tapia-2} indicated that $\alpha_e$ is the order of $1$ and can be either positive or negative contingent upon the nature of the cell boundary. Further, the authors have observed a good agreement between the experimental and theoretical studies for values of $\alpha_e$ ranging from $-1.0$ to $1.47$. Later, Francisco et al.\cite{Francisco-1,Francisco-2} have derived an approximate expression of the stress jump parameter and correlated the expression with the existing experimental results; concluding that $\alpha_e \sim 0-1.3$. Therefore, one can reasonably choose the values of $\alpha_e$ in the range $\mid \alpha_e \mid \lesssim 1$. In addition to the stress conditions, in the frame of the reference squirmer, the velocity field in the effective Brinkman phase is expected to agree with the migration velocity of the squirmer in the limit $r\to \infty$. Consequently, we have
\begin{equation}
\mathbf{v}^e \to \mathbf{U} \quad \text{as} \quad r\to \infty. \label{bc-6-ch7}
\end{equation}
It may be noted that the effective medium model entails the assumption that the squirmers are homogeneously distributed across the medium. Therefore, without the loss of generality, $\mathbf{U}$ indicates the locomotion speed of any arbitrary squirmer packed in the ensemble. Theoretical studies have suggested that homogeneous ensembles with identical mobility coefficients can improve the self-assembly among the constituents \citep{Hocking} and can give rise to stable capsid-like metastable structures at a given instance of time \citep{Mallory}. These are reasonably similar to the cell model-like arrangement of the squirmers in the present work. We further assume that the squirmers generate axisymmetric forcing on the fluid (i.e., $\mathbf{v}_s$ is axisymmetric) and consequently move with a velocity $\mathbf{U}=U \, \mathbf{\hat{k}}$ that is to be determined, where $\mathbf{\hat{k}}$ is the unit vector along the $z$-axis.

\subsection{Analytical solution of the hydrodynamic problem}
\label{solution-ch7}

\begin{figure}
\centering
\includegraphics[scale=0.4]{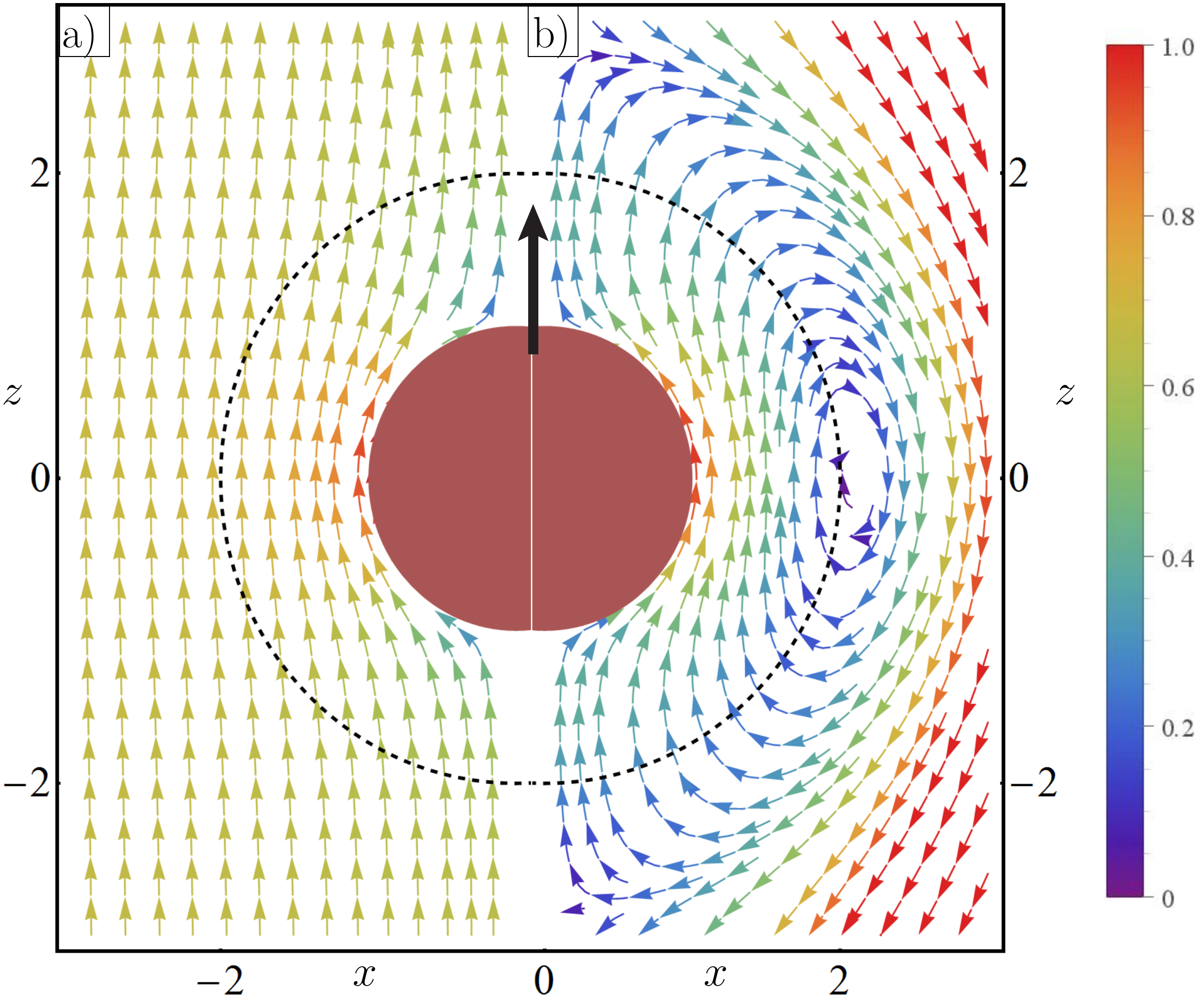}
\caption{\label{Figstream-ch7} The stream-lines ($x-z$ plane) of the velocity field of $a)$ an isolated squirmer in a unbounded medium and $b)$ a squirmer in presence of neighbouring squirmers in an effective Brinkman medium. Here, we have fixed $j=2$ as the position of the cell boundary while the other parameters were fixed as $\beta=1.0$, $\chi_e=1.0$, and $\alpha_e=1.0$. One may note that, in case of effective medium, the deviation in the streamlines starts occurring adjacent to the position of the cell boundary.}
\end{figure}

The governing equations (eqs. (\ref{gov-1-ch7}), (\ref{gov-2-ch7})) in both the fluid phases subjected to the aforementioned interface and boundary conditions (eqs. (\ref{bc-1-ch7})-(\ref{bc-6-ch7})) are exactly solvable. Owing to the axisymmetry of the problem, one can always express the velocity flow fields in terms of the associated stream functions which have been extensively used in order to solve such problems \cite{Happel}. However, it has been shown that for spherical geometry, the velocity and pressure fields expressed in terms of two scalar functions that are biharmonic and harmonic in nature can be treated as an alternative efficient tool to solve such problems \cite{Padmavathi-PM, GPRS, Padmavathi-QJ} . For the present model, we adopt this technique and manifest the velocity and pressure fields as
\begin{align}
\mathbf{v} & = \nabla \times \nabla \times (A^i \mathbf{r}) + \nabla \times (B^i \mathbf{r}), \\
p & = p_0 + \frac{\partial}{\partial r}[ r (\nabla^2 - \chi_e^2) A^i],
\end{align}
where $A^i(r,\theta,\phi)$ and $B^i(r,\theta,\phi)$ are the scalar functions that satisfy the following equations: $\nabla^2(\nabla^2 -\chi_e^2)\, A^i=0$ and $(\nabla^2 -\chi_e^2)\,B^i=0$, and $i$ can be $en$ or $e$ depending on whether the flow is within the cell envelope or in Brinkman region. Here, $\mathbf{r}$ is the position vector and $p_0$ is the constant pressure. It may be noted that the non-zero representations of the parameter $\chi_e$ indicates the flow in the Brinkman region (i.e., $i=e$), whereas $\chi_e=0$ represents the flow field in the clear fluid or Stokes region (i.e., $i=en$). In our current system of interest, the axisymmetric migration of the squirmers annihilates the rotation degrees of freedom, subsequently, we have $B^i$ becoming zero throughout the domain of interest. Further, the solution of $A^i$ (for both $i=en, \, e$) can be readily obtained in spherical polar coordinates (see Appendix~\ref{App-A}). Therefore, the above-mentioned interface conditions (eqs. \ref{bc-1-ch7}-\ref{bc-5-ch7}) in terms of scalar $A^i$ reduce to a system of linear equations with the unknown constants as variables. Correspondingly, the unknown coefficients present in the scalar and the associated velocity fields can be appropriately obtained in a straightforward way. Fig.~\ref{Figstream-ch7} compares the flow fields for an unbounded squirmer (in pannel $a)$) with a squirmer in a suspension modeled as an effective Brinkman medium (in pannel $b)$). The unbounded squirmer swims in a uniform axisymmetric streamline path, whereas, it is observed that in the presence of nearby squirmers, the streamlines start deviating adjacent to the cellular boundary and an additional closed-loop vortex is formed on the adjacent sides of the squirmer (where the magnitude of the squirmer slip is maximum). Interestingly, the stress jump condition (as in eq. (\ref{bc-5-ch7})) comprises both the tangential velocity (i.e., the slip distribution) and the stress parameter $\alpha_e$. However, as the prescribed stress is isotropic in nature, the position of this emerging vortex is defined by the slip distribution which is maximum on the adjacent sides of the squirmers.

\subsection{Force acting on the surface of the squirmer}
\label{Force-ch7}

For evaluation of the drag on the surface of the squirmer, we proceed with computing the stress components $\bm{\Sigma}^{en}$ of the reference squirmer inside the cell envelope and evaluate the drag $\mathbf{D}= \int \bm{\Sigma}^{en} \cdot \mathbf{\hat{n}} \, ds$, where $ds$ is the elementary surface element of the squirmer, which can be resolved as
\begin{equation}
\mathbf{D}= \mathbf{D}_H + \mathbf{D}_A, \label{drag-ch7}
\end{equation}
where the subscripts $H$ and $A$ are used to denote the rigid body hydrodynamic drag and active forcing on the surface of the reference squirmer, respectively. Further, for axisymmetric swimming, the magnitudes non-zero component of the drag forces $\mathbf{D}_H$ and $\mathbf{D}_A$ are in the $z$-direction and can be represented as
\begin{equation}
D_H = \frac{D_1}{D_2}, \quad D_A = \frac{D_3}{D_2} \label{HA-drag-ch7},
\end{equation}
where the detailed structure of $D_1= D_1(U, \chi_e, j, \alpha_e)$, $D_2= D_2(\chi_e,j, \alpha_e)$, and $D_3= D_3(\beta, \chi_e, \alpha_e, j)$ are given in the Appendix~\ref{App-C}. Now, in the view of macroscopic Darcy medium, the overall bed permeability constituted by the squirmers relates the average far-field velocity of the suspension $\langle \mathbf{U} \rangle$ and the macroscopic pressure gradient ($P^e$) are related in accordance to Darcy's law as, 
\begin{equation}
\label{darcy-law-ch7}
\langle \mathbf{U} \rangle = - \frac{k_e}{\mu} \nabla P^e,
\end{equation}
where in our case the macroscopic pressure term $P^e$ consists both the effective pressure of the consituent particles ($p^e$)  and the constant pressure gradient ($\nabla p^f$) applied to the system, i.e., $P^e = p^e + (\nabla p^f \cdot \mathbf{r})$. Through the effective medium approximation we have already evaluated the drag force per squirmer, further the pressure in the effective medium balances the total drag force which is force per particle multiplied by the number of squirmer per unit volume as
\begin{equation}
\label{pressure-gradient-ch7}
-\nabla p^e= \frac{\Phi \mathbf{D}}{4 \pi a^3/3},
\end{equation}
where $\mathbf{D}$ is the force on a single squirmer (eq. (\ref{drag-ch7})) and $\Phi=(1-\Xi)=1/j^3$ is the volume fraction. Hence from eqs. (\ref{darcy-law-ch7}) and (\ref{pressure-gradient-ch7}) we have
\begin{equation}
\label{chi_e-phi-relation-ch7}
\mathbf{D}= 6 \pi \mu a \langle \mathbf{U} \rangle \left(\frac{2 a^2}{9 \Phi k_e} \right) + \frac{ 4 \pi a^3}{3 \Phi} \nabla p^f.  
\end{equation}
Henceforth, by utilizing the components of the drag and eliminating the migration velocity $\mathbf{U}$, we can establish a relationship between the parameters of the effective Brinkman medium $\chi_e= a^2/k_e$ and $\Phi$. Addressing the difficulty of giving a compact expression, we attempt to obtain an empirical relation by satisfying the identity established in eq. (\ref{chi_e-phi-relation-ch7}).

\subsection{Correlation between $\chi_e$ and $\Phi$}

\begin{figure}
\centering
\includegraphics[scale=1.0]{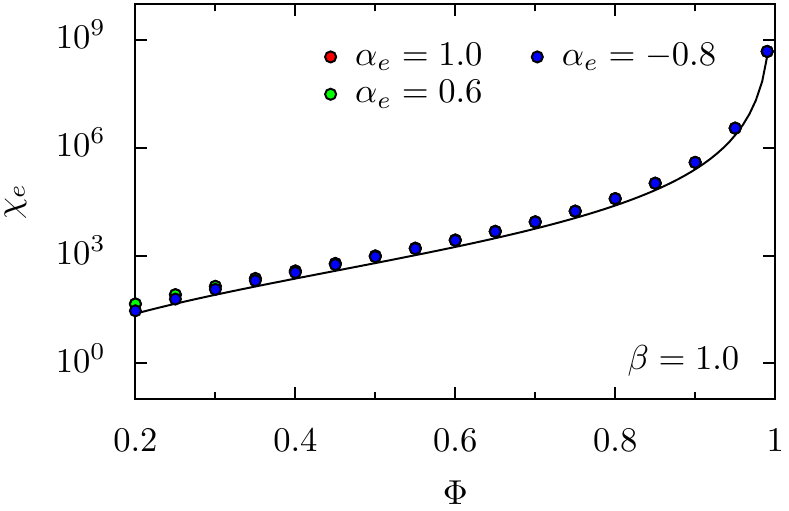}
\caption{\label{Fig7-ch7} The $\chi_e$ and $\Phi$ variation for $\beta= 1.0$. We note that the points for different $\alpha_e$ nearly coincide. The solid line indicates the best-fitted curve (in eq. (\ref{CK-relation-ch7})) which is averaged over the range $\mid \alpha_e \mid \leq 1$.}
\end{figure}

We recall that in case of bed of passive rigid particles, the classical hydraulic radius theory of Carman-Kozeny leads to the relationship $\chi_e \simeq 45\, \Phi^2/(1- \Phi)^3$ \citep{Sharanya-16, Carman-Book, Anton}. In the view of a suspension of active squirmers, we initiate a more generic relation given in the form
\begin{equation}
\label{fitting-function-ch7}
\chi_e= \mathcal{A}\, \frac{\Phi^{\varepsilon_1}}{(1- \Phi^{\varepsilon_2})^{\varepsilon_3}},
\end{equation}
\begin{table*}[ht]
\begin{center}
\begin{tabular}{||p{10mm}|p{10mm}|p{10mm}| p{10mm}|p{10mm}||}
\hline
$\beta $ & $\mathcal{A} $ & $\varepsilon_1$ & $\varepsilon_2$ & $\varepsilon_3$ \\ [0.5ex]
\hline\hline
0.5 & 281 & 2 & 1 & 3 \\
\hline
1 & 308 & 2 & 1 & 3 \\
\hline
1.5 & 376 & 2 & 1 & 3 \\
\hline
2 & 613 & 2 & 1 & 3 \\ [1ex]
\hline
\end{tabular}
\end{center}
\caption{\label{Table-ch7} The stress averaged best-fitted parameters for different slip coefficients $\beta$ (with respect to the velocity scale $U$)}
\end{table*}
where the parameters $\mathcal{A},\,\varepsilon_1,\,\varepsilon_2,\,$ and $\varepsilon_3$ are obtained by fitting eq. (\ref{fitting-function-ch7}) to the values extracted from the relation established in eq. (\ref{chi_e-phi-relation-ch7}). Surprisingly, the best-fitted coefficients remains unaltered from the Carman-Kozeny values as we have $\varepsilon_1 \sim 2.0$, $\varepsilon_2 \sim 1.0$, and $\varepsilon_3 \sim 3.0$. It may be noted that throughout the calculation we have kept the term containing external pressure gradient constant as $\mid \nabla p^f \mid = 1$ with respect to the velocity scale $U$. However, we notice that the prefactor $\mathcal{A}$ which contains information about the net surface area, hydraulic radius and the systems characteristics depends only on the intrinsic activity of the squirmer ($\beta$) and the stress jump coefficient ($\alpha_e$) prescribed across the cell boundary. In particular, we notice that $\mathcal{A}$ modifies meagerly with $\alpha_e$ in the given range $\mid \alpha_e \mid \leq 1$ (see fig~\ref{Fig7-ch7}), but increases significantly with the activity parameter $\beta$. In table~\ref{Table-ch7}, we classify the stress averaged best-fitted parameters for different $\beta$ values (considered with respect to the velocity scale $U$) which indicates a monotonically increasing trend. This indicates that in active suspensions of squirmers with higher slip coefficients the resistance experienced by a reference squirmer is significantly higher. However, in the present study, we fix the value $\beta=1$ and proceed with a Carman-Kozeny like correlation for an active suspension represented by the following equation
\begin{equation}
\chi_e \simeq 308\, \frac{\Phi^2}{(1- \Phi)^3}. \label{CK-relation-ch7}
\end{equation}

\section{Swimming problem}

In an unbounded fluid, a squirmer is allowed to move freely with a locomotion speed $\mathbf{U}_s= 2\beta/3\,\, \mathbf{\hat{k}}$. Similarly, to estimate how the neighboring squirmers restrain the autonomous swimming and lead to different emergent swimming speeds for a single squirmer, we must resort to the force balance condition. For any reference squirmer, present in the suspension, in the absence of any external forces, the aforementioned hydrodynamic and active forces (in eq. (\ref{HA-drag-ch7})) must balance and satisfy the condition $\mathbf{D}=0$. Correspondingly, the migration velocity is determined as $\mathbf{U}= U_1/U_2\, \mathbf{\hat{k}}$ where
\begin{align}
U_1 & = \left( 3 \chi_e \alpha_e + 3 j \chi_e^2 (1+ \alpha_e) + 2 j^7 \chi_e^3 (1+\alpha_e) - 5 j^3 \chi_e^2 (3 + \alpha_e) + 2j^6 \chi_e^2 (6 + \alpha_e) \right. \nonumber \\ & \left. + \, 6 j^5 \chi_e (15 + 7 \alpha_e) - 5 j^4 (-18 + \chi_e^3 (1+\alpha_e)) + 3 j^2 (5 \chi_e \alpha_e + \chi_e^3 (1+\alpha_e)) \right) \beta \label{mig-v-1} \\\\
U_2 & = 3j (45 j^3 + 45 j^4 \chi_e + (-1 - 5 j^3 + 21j^3) \chi_e^2 + j(-1 - 5 j^3 + 6 j^5)\chi_e^3) \nonumber \\ & + 3 (-1 - 5 j^3 + 6 j^5) \chi_e (1+ j \chi_e (j+ \chi_e)) \alpha_e. \label{mig-v-2}
\end{align}
As a first check, we recover the strength of the squirming velocity of a standard squirmer $U=U_s=2\beta/3$ in the limit $\chi_e \to 0$, which corresponds to the case when the targeted squirmer is moving freely without any resistance from the ensemble. This can be taken as a reference scale while analyzing the squirmer's migration velocity and other relevant characteristics. It may be noted that, while $\chi_e$ signifies the effective presence of other squirmers in the suspension, $j^3=1/\Phi$ controls the packing fraction or the interaction range between the squirmers. Under the assumption of a homogeneous suspension an empirical relation in eq. (\ref{CK-relation-ch7}) appropriately relates these two morphological parameters. In what follows, we employ this and allude to the effectiveness of squirmers' network in terms of its packing or volume fraction $\Phi$ and explore its consequences on the relevant swimming characteristics. Note that one can effectively use the reciprocal theorem \cite{Stone-Sam} to bypass the calculations of the flow fields and obtain the swimming speed. However, information such as power dissipation, swimming efficiency, and effective viscosity cannot be obtained without knowing the flow around the targeted squirmer.

\section{Results}

\subsection{Migration velocity}

\begin{figure}
\centering
\includegraphics[scale=1.0]{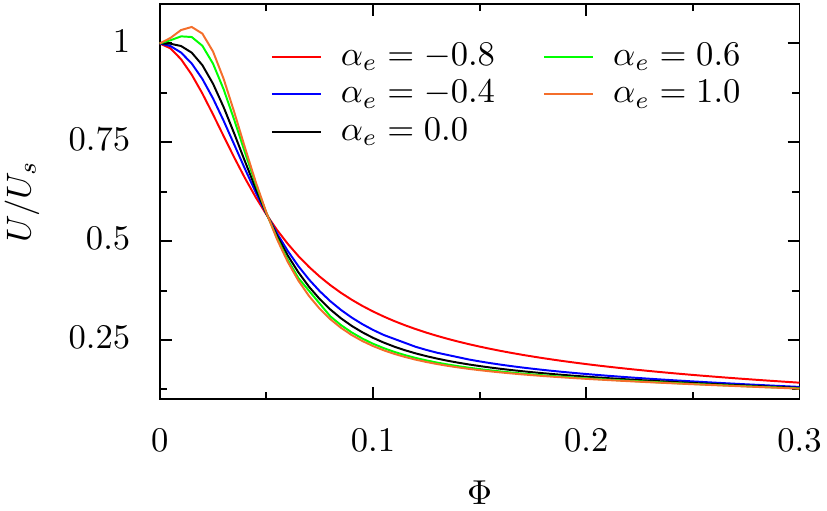}
\caption{\label{Fig2-ch7} The variation of the scaled migration velocity $U/U_s$ with the volume fraction $\Phi$ of the effective Brinkman medium at various values of the interactive stress $\alpha_e$, the migration velocity exhibits a maximum at $\Phi \sim 0.02$ for $\alpha_e>0$. }
\end{figure}

One may note that the underlying assumptions of the cell model and the effective medium model are valid for dilute to an intermediate concentration range of a suspension. Therefore, one may identify many intriguing behaviors of the targeted squirmer by varying $\Phi$ in the range $\Phi \lesssim 0.3$. Figure~\ref{Fig2-ch7} a) depicts that, for initial values of $\Phi$, the swimming velocity initially increases and attains a maximum close to $\Phi \sim 0.02$ for $\alpha_e>0$. Earlier, exact numerical studies for a lesser number of squirmers have indicated a similar development of a higher migration velocity of squirmer in the presence of a nearby squirmer \cite{Ishikawa-JFM,Ishikawa-2,Papavassiliou}. The $\Phi\,(= (a/b)^3 \sim 0.02)$ value at which the relative maximum ($\Delta U \sim 4 \% $) occurs indicates that the nearest neighbouring squirmers are located at a distance $b \sim 3.67\, a$. Further, we note that the higher concentration of squirmers ($\Phi \gg 0.02$) increases the targeted swimmer's resistance, and subsequently, the locomotion speed decreases. Note that there exists an additional $\Phi$ ($\sim 0.05 $) value, where the curves for different $\alpha_e$ traverse each other. Beyond that, the deployment with the maximum $\alpha_e$ decays the fastest although it yields the foremost maximum value. Therefore, $\alpha_e$ emerges as a key parameter that quantifies the magnitude of increase or decrease compared to a standard free squirmer. At initial values of $\Phi(<0.02)$, the alliance and a net positive interaction ($\alpha_e>0$) among the squirmers prompts an increase in the swimming velocity. However, towards higher $\Phi \,(> 0.05)$ values, the squirmer eventually gets trapped due to higher resistivity, and stronger interactive stress will reduce swimming velocity in the fastest way. Moreover, from the expression of $U$, we identify that in the limit $\Phi \to 1$, $U$ asymptotically diminishes. Similar behavior is observed in the case of a single squirmer in a passive Brinkman medium \cite{Nganguia}.

\subsection{Power dissipation and efficiency}
\label{Power-efficiency-ch7}

To assess the cost of having higher motility at certain instances, it is necessary to consider the power ($P$) expended during the swimming process and examine the corresponding swimming efficiency. Since the work done by the squirming motion is equal to the power dissipation in the fluid, we have $P=-\int \bm{\Sigma}^{en} \cdot \mathbf{\hat{n}} \cdot \mathbf{v}^{en} \, \, ds$. The integral can be evaluated with the help of constitutive relations and the corresponding flow fields of the swimmer. The expression for power dissipation for an autonomous squirmer was calculated by Lighthill \cite{Lighthill} and Blake \cite{Blake} with the first two modes of the tangential velocity actuation. However, within our simplified framework, we calculate the leading order contribution (with the first mode) of the power dissipation of a single squirmer in an assemblage (see Appendix~\ref{App-C}) which agrees with the power dissipation of a standard squirmer $P=P_s=16 \pi \mu\, \beta^2/3$ in the limit $\Phi \to 0$.

\begin{figure}
\includegraphics[scale=0.9]{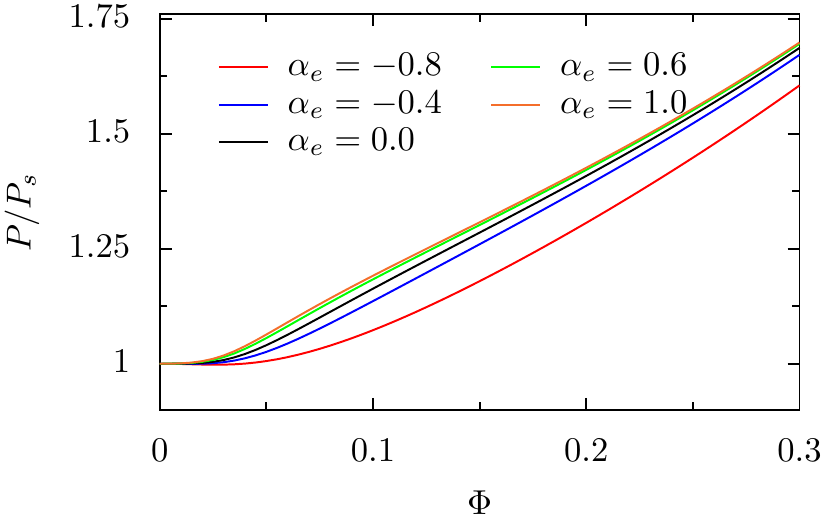}
\includegraphics[scale=0.9]{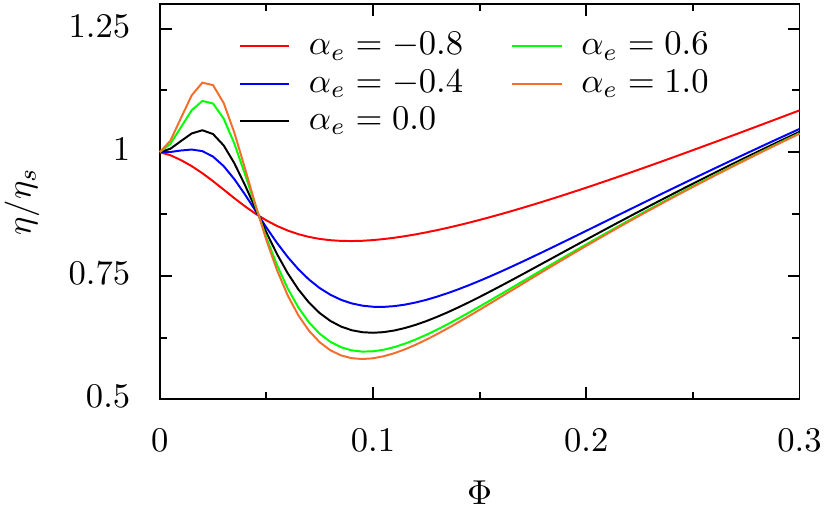}
\caption{\label{Fig3-ch7} The variation of dissipative power ($P$) and swimming efficiency ($\eta$) with the volume fraction $\Phi$ for different $\alpha_e$ values. $P_s$ and $\eta_s$ denote the power dissipated and the efficiency of a free autonomous squirmer.}
\end{figure}

Figure~\ref{Fig3-ch7} depicts that for increasing packing fraction ($\Phi$), $P$ exhibits a monotonically increasing trend. The augmentation being the highest for the specific arrangements, which lead to a higher locomotion speed (see figs~\ref{Fig2} and ~\ref{Fig3}). The behavior of $P$ simply remarks that the neighboring squirmers and the associated interactions increase the targeted squirmer's energy expenditure, which is insufficient to estimate the corresponding swimming efficiency. To further investigate, we probe the dissipated power $P$ to the power output $D_{H}\,U$ and calculate the hydrodynamic efficiency $\eta=D_{H} U/P$. The behavior of $\eta$ with $\Phi$ is depicted in fig~\ref{Fig3-ch7}. It is observed that on increasing $\Phi$, $\eta$ varies non-monotonically with a maximum around $\Phi\sim 0.02 $ (for $\alpha_e \geq 0$), which yields the maximum $U$. However, the relative increment in the efficiency ($\Delta \eta \sim 14 \%$) is higher than the increase in the case of $U$. Typically, $\eta$ compares the total power dissipation to the power against the medium's net hydrodynamic drag. At lower $\Phi$, the power output of the medium scales as $D_H U \sim (1 + \Phi^3)$ which is higher than the dissipative power scaling as $P \sim (1+ \Phi^{7/3}) $. Hence, the relative increase in $P$ lags the net power output of the squirmer. This persuades the maximization of swimming efficiency around the initial values of $\Phi$. Further, similar to $U$, at $\Phi \sim 0.05$, the efficiencies for different $\alpha_e$ values cross over each other. Therefore, $\Phi \sim 0.05$ marks the onset of a higher resistive regime of the squirmer concerning both the migration velocity and swimming efficiency.

\bibliography{Dhar_Langmuir}

\section{Appendix}

\subsection{Solution of Brinkman and Stokes equations}
\label{App-A}

A complete general solution of the Brinkman equation (as in eq.~\ref{gov-2}) in the exterior phase ($r>b/a$) can be expressed in terms of two scalar functions as
\begin{align}
\mathbf{v^e} & = \nabla \times \nabla \times (A^e \mathbf{r} ) + \nabla \times (B^e \mathbf{r}), \label{ve} \\
p^e &= p_{0} + \frac{\partial}{\partial r}[ r (\nabla^2 - \chi_e^2) A^e], \label{pe}
\end{align}
where $A^e(r,\theta,\phi)$ and $B^e(r,\theta,\phi)$ are two scalar functions that satisfy the equations $\nabla^2(\nabla^2 - \chi_e^2) A^e=0$ and $(\nabla^2 - \chi_e^2) B^e=0$, respectively, and $\chi_e=a^2/k_e$ is the dimensionless resistivity of the effective Brinkman medium with porosity $k_e$. The general series solution of aforementioned governing equations in $(r,\theta,\phi)$ spherical polar coordinate system can be written as \cite{Padmavathi-PM,JPrakash},
\begin{align}
A^e & = \sum_{n=1}^{\infty} \left( a_{n}^e r^{n} + b_{n}^e u_n(\chi_e r) + \frac{c_{n}^e}{r^{n+1}} + d_{n}^e v_n(\chi_e r) \right) S_{n}(\theta,\phi), \label{sc-1} \\
B^e & = \sum_{n=1}^{\infty} \left( e_{n}^e u_n(\chi_e r) + f_{n}^e v_n(\chi_e r) \right) H_{n}(\theta,\phi), \label{sc-2}
\end{align}
where $a^e_{n},b^e_{n},c^e_{n},d^e_{n},e^e_{n},f^e_{n}$ are the arbitrary coefficients and $u_n(\chi_e r)$ and $v_n(\chi_e r)$ are modified spherical Bessel functions of the first and second kind, respectively. The functions $S_{n}(\theta,\phi)$ and $H_{n}(\theta,\phi)$ are the spherical harmonics, defined as $S_{n}(\theta,\phi)= \sum_{m=0}^{n} P_{n}^{m}(\cos\theta)(A_{nm}\cos(m\phi) + B_{nm}\sin(m\phi))$ and $H_{n}(\theta,\phi)=\sum_{m=0}^{n}  P_{n}^{m}(\cos\theta)(C_{nm}\cos(m\phi) + D_{nm}\sin(m\phi))$, where the coefficients $A_{nm}, B_{nm}, C_{nm}, D_{nm}$ are known. 

We further assume that the scalar functions corresponding to the ambient velocity and the pressure fields, which are solutions of the Brinkman equation as $\mathbf{v_{0}}$ and $p_{0}$, respectively, are given by $A_{0}$ and $B_{0}$. The ambient velocity is significant in the limit $r\to\infty$, therefore, the spatially decaying terms in eqs.~(\ref{sc-1}) and (\ref{sc-2}) disappear. Thus, $A_{0}$ and $B_0$ are given by
\begin{align}
A_{0} & =\sum_{n=1}^{\infty} \Big(a_{n}^e r^{n} + b_{n}^e u_n(\chi_e r) \Big)S_{n}(\theta,\phi), \label{am1} \\
B_{0}& = \sum_{n=1}^{\infty}  e_{n}^e u_n(\chi_e r) H_{n}(\theta,\phi), \label{am2}
\end{align}
where  $a_{n}^e,b_{n}^e,e_{n}^e$ are known and can be the extracted for a given ambient velocity field. For uniform axisymmetric migration in the $z$ direction, i.e., $\mathbf{v_{0}}=U\,\mathbf{\hat{k}}$, the corresponding scalars $A_{0}$ and $B_0$ can be extracted using eqs. (\ref{ve}) and (\ref{am1}) as
\begin{equation}
A_{0}= \frac{U r \cos\theta}{2}, \quad B_0=0, 
\end{equation}
where $U$ is the unknown far field migration velocity of the squirmer. Correspondingly, we have $a_1^e= U/2$, and $a_n^e=0$ for $n\geq 2$ with all the other coefficients ($b_n^e$, $e_n^e$) being invariably zero. 

On the other hand, for the resulting Stokes flow (as in eq~(\ref{gov-1})) inside the envelope ($1<r<b/a$), we have the following relations for the velocity and pressure fields,
\begin{align}
\mathbf{v^{en}} & = \nabla \times \nabla \times (A^{en} \mathbf{r} ) + \nabla \times (B^{en} \mathbf{r}), \label{ven} \\
p^{en} &= p_{0} + \frac{\partial}{\partial r}[ r \nabla^2 A^{en}], \label{pen}
\end{align}
where $A^{en}(r,\theta,\phi)$ and $B^{en}(r,\theta,\phi)$ satisfy the equations $\nabla^4 A^{en}=0$ and $\nabla^2 B^{en}=0$, respectively, with general solutions,
\begin{align}
A^{en} & = \sum_{n=1}^{\infty} \left( a_{n}^{en} r^{n} + b_{n}^{en} r^{n+2} + \frac{c_{n}^{en}}{r^{n+1}} + \frac{d_{n}^{en}}{r^{n-1}} \right) S_{n}(\theta,\phi), \label{sc-3} \\
B^{en} & = \sum_{n=1}^{\infty} \left( e_{n}^{en} r^{n} + \frac{f_{n}^{en}}{r^{n+1}} \right) H_{n}(\theta,\phi), \label{sc-4}
\end{align}
where $a^{en}_{n},b^{en}_{n},c^{en}_{n}, d^{en}_{n}, e^{en}_{n}, f^{en}_{n}$ are the arbitrary unknown coefficients. This summaries the calculation of the scalar fields ($A^{e}, B^{e}$) and ($A^{en}$, $B^{en}$) require to compute the velocity and pressure fields in the exterior ($r>b/a$) and interior ($1<r<b/a$) phases of the effective Brinkman medium. Further, for our axisymmetric effective Brinkman medium, we have  the $B^e$ and $B^{en}$ becoming zero throughout the domain of interest. Therefore, the remaining unknown coefficients ($c_n^e$, $d_n^e$) and ($a^{en}_{n},b^{en}_{n},c^{en}_{n}, d^{en}_{n}$) are to be obtained from the suitable boundary conditions mentioned in section~\ref{math-model} (eqs. ~(\ref{bc-1})-(\ref{bc-5})).

\subsection{The coefficients}
\label{App-C}

The drag coefficients of the squirmer calculated in eq. (\ref{HA-drag}) are evaluated as
\begin{align}
D_1 & = 8 \pi \mu \left( - 3 U \chi_e \alpha_e - 15 j^3 U \chi_e \alpha_e - 3 j U \chi_e^2 (1+ \alpha_e) - 3 j^2 U \chi_e^2 (1+ \alpha_e) + 18 j^7 U \chi_e^3 (1+\alpha_e) \right. \nonumber \\ & \left. + 9 j^6 U \chi_e^2 (7 + 2 \alpha_e) - 15 j^4 U (-9 + \chi_e^2 (1+\alpha_e)) - 3 j^5 U \chi_e (-45 - 6 \alpha_e + 5 \chi_e^2 (1+\alpha_e)) \right) \\
D_2 & = 180 j^4 + 180(-1+j^4) \chi_e + 3(-1 +j)^3 j (9 + 8 j) \chi_e^2 + (-1 + j)^4 j (4 + j(7 + 4j)) \chi_e^3 \nonumber \\ & +\chi_e \alpha_e (-9 - 30 j^2 -45 j^4 + 84j^5 + (-1+j)^4 (4 + j(7 +4j))\chi_e (1 + \chi_e j) \\
D_3 & = 8 \pi \mu \left((-3 \alpha_e \chi_e - 3 j \chi_e^2 (1+\alpha_e) - 2 j^7 \chi_e^3 (1+\alpha_e) + 5 j^3 \chi_e^2 (3 + \alpha_e) - 2 j^6 \chi_e^2 (6 + \alpha_e) \right. \nonumber \\ & \left. - 6 j^5 \chi_e (15 + 7 \alpha_e) - 3 j^2 \chi_e (\chi_e^2 + (5 + \chi_e^2) \alpha_e) + 5 j^4 (-18 + \chi_e^3 (1+\alpha_e)) \right) \beta 
\end{align}

The expression calculated for power dissipation can be written as $P= P_1/P_2$ where $P_1$ and $P_2$ are given by
\begin{align}
P_1 & = 8 \pi \mu \left( 30 \Phi + \Phi^{7/3} \chi_e \alpha_e + (1+ \alpha_e)( \Phi^2 \chi_e^2  + 4 \chi_e^3 + \Phi^{5/3} \chi_e^3) + 2 \Phi^{1/3} \chi_e^2 (7 + 2 \alpha_e) \right. \nonumber \\ & \left. + 2 \Phi^{2/3} \chi_e (15 + 2 \alpha_e) \right) \beta^2, \\
P_2 & =  \left( -45 \Phi - 45 \Phi^{2/3} \chi_e - 21 \Phi^{1/3} \chi_e^2 + 5 \Phi \chi_e^2 + \Phi^2 \chi_e^2 - 6 \chi_e^3 + 5 \Phi^{2/3} \chi_e^3 + \Phi^{5/3} \chi_e^3 \right. \nonumber \\  & \left. + \chi_e \alpha_e (-6 + \Phi^{2/3} (5 + \Phi) ) (\Phi^{2/3} + \Phi^{1/3} \chi_e + \chi_e^2) \right). 
\end{align}  
Further the derived relation between the $\chi_e$ and $\Phi$ in eq. (\ref{CK-relation}) can be substituted to simply the expression in terms of $\Phi$. 

The calculated viscosity function $f(\Phi, \alpha_e, \beta)=f_1/f_2$ (in eq. (\ref{eff-viscosity})) which dictates the change in the effective viscosity of the suspension is given by,

\begin{align}
f_1 & = \left( 90 \beta \Phi^{1/3} + 90 \beta \Phi^{2/3} + 90 \beta \Phi - 675 \beta \Phi^{4/3} - 675 \beta \Phi^{5/3} + 27045 \beta \Phi + 298880 \beta \Phi^{7/3} \right. \nonumber \\ & \left. + 29880 \beta \Phi^{8/3} - 122580 \beta \Phi^3 \right), \\
f_2 & = \left( 360 + 360 \Phi^{1/3} + 360 \Phi^{2/3} - 2800 \Phi - 2800 \beta \Phi^{4/3} + (10800 + 14784 \alpha_e) \Phi^{5/3} \right. \nonumber \\ & \left. + (120980 + 14784 \alpha_e) \Phi^2  + (12960 + 2469 \alpha_e) \Phi^{7/3} - (544320 + 86240 \alpha_e) \Phi^{8/3} \right. \nonumber \\ & \left. - (574560 + 86240 \alpha_e) \Phi^{3} \right).  
\end{align}

\end{document}